\documentclass[12pt]{article} 

\usepackage[utf8]{inputenc}
\usepackage[T1]{fontenc} 
\usepackage[english]{babel}           

\usepackage[pdftex]{graphicx}
\usepackage{epsfig}
\usepackage{graphicx}
\usepackage{comment}
\usepackage{latexsym}
\usepackage{hyperref}
\usepackage{amsmath}
\usepackage{mathrsfs}
\usepackage[usenames, dvipsnames]{color}
\usepackage{amsbsy}
\usepackage{amssymb}
\usepackage{amsthm}
\usepackage{amsfonts}
\usepackage{cite}
\usepackage{enumitem}
\usepackage{xcolor}
\usepackage{color}
\usepackage{mathtools}
\usepackage{caption} 
\usepackage{subcaption} 

\usepackage{diagbox}

\usepackage[normalem]{ulem}

\newcommand{\be}[0]{\begin{equation}}
\newcommand{\ee}[0]{\end{equation}}

\renewcommand{\thefootnote}{\fnsymbol{footnote}}

\newcommand{\Z}{\mathbb{Z}}

\renewcommand{\O}{{\cal O}}

\renewcommand{\d}{{\rm d}}

\newcommand{\where}{\mbox{where}}
\newcommand{\with}{\mbox{with}}

\renewcommand{\and}{\mbox{and}}

\newcommand{\espD}{\phantom{\!\!\underset{\displaystyle |}{\cdot}}}

\newcommand{\bm}{\boldmath} 

\newcommand{\white}{\color{white}}

\newcommand{\N}{{\cal N}}
\newcommand{\F}{{\cal F}}

\newcommand{\goh}{\mathfrak{h}}
\newcommand{\gog}{\mathfrak{g}}

\newcommand{\Vone}{V_{\mbox{\scriptsize 1-loop}}}
\newcommand{\x}{$-$}
\newcommand{\px}{$+$}

\catcode`\@=11
\def\marginnote#1{}
\newcount\hour
\newcount\minute
\newtoks\amorpm
\hour=\time\divide\hour by60 \minute=\time{\multiply\hour by60
\global\advance\minute by-\hour}
\edef\standardtime{{\ifnum\hour<12 \global\amorpm={am}%
        \else\global\amorpm={pm}\advance\hour by-12 \fi
        \ifnum\hour=0 \hour=12 \fi
        \number\hour:\ifnum\minute<10 0\fi\number\minute\the\amorpm}}
\edef\militarytime{\number\hour:\ifnum\minute<10 0\fi\number\minute}
\def\draftlabel#1{{\@bsphack\if@filesw {\let\thepage\relax
   \xdef\@gtempa{\write\@auxout{\string
      \newlabel{#1}{{\@currentlabel}{\thepage}}}}}\@gtempa
   \if@nobreak \ifvmode\nobreak\fi\fi\fi\@esphack}
        \gdef\@eqnlabel{#1}}
\def\@eqnlabel{}
\def\@vacuum{}
\def\draftmarginnote#1{\marginpar{\raggedright\scriptsize\tt#1}}
\def\draft{\oddsidemargin -.2truein
        \def\@oddfoot{\sl preliminary draft \hfil
        \rm\thepage\hfil\sl\today\quad\militarytime}
        \let\@evenfoot\@oddfoot \overfullrule 3pt
        \let\label=\draftlabel
        \let\marginnote=\draftmarginnote
   \def\@eqnnum{(\theequation)\rlap{\kern\marginparsep\tt\@eqnlabel}%
\global\let\@eqnlabel\@vacuum}  }
\def\thebibliography#1{
\vskip 0.5cm \centerline{\bf \Large References}
\list{
[\arabic{enumi}]}{\settowidth\labelwidth{[#1]}
\leftmargin\labelwidth
\advance\leftmargin\labelsep
\usecounter{enumi}}
\def\newblock{\hskip .11em plus .33em minus .07em}
\sloppy\clubpenalty4000\widowpenalty4000
\sfcode`\.=1000\relax}

\renewcommand{\theequation}{\arabic{section}.\arabic{equation}}
\renewcommand{\section}{\setcounter{equation}{0}\@startsection
{section}{1}{0mm}{-\baselineskip}{0.5\baselineskip} {\normalfont\Large\bfseries}}
\renewcommand{\subsection}{\@startsection
{subsection}{2}{0mm}{-\baselineskip}{0.5\baselineskip} {\normalfont\large\bfseries}}
\renewcommand{\subsubsection}{\@startsection
{subsubsection}{3}{0mm}{-\baselineskip}{0.5\baselineskip}
{\normalfont\normalsize\slshape}}

\begin{document}


\begin{titlepage}
\begin{flushright}
CPHT-RR105.122019, December 2019
\vspace{1cm}
\end{flushright}
\begin{centering}
{\bm\bf \Large HETEROTIC $-$ TYPE I DUAL PAIRS,  \\ \vspace{0.2cm}
RIGID BRANES AND BROKEN SUSY}

\vspace{5mm}

 {\bf Carlo Angelantonj$^{1}$, Herv\'e Partouche$^2$ and Gianfranco Pradisi$^3$}

 \vspace{3mm}

$^1$ Dipartimento di Fisica, Universit\`a di Torino, \\ INFN Sezione di Torino and Arnold-Regge Center\\
Via P. Giuria 1, 10125 Torino, Italy\\
{\em  carlo.angelantonj@unito.it}

$^2$  {CPHT, CNRS, Ecole polytechnique, IP Paris, \\F-91128 Palaiseau, France\\
{\em herve.partouche@polytechnique.edu}}

$^3$ Dipartimento di Fisica, Universit\`a di Roma Tor Vergata,\\  
INFN, Sezione di Roma ``Tor Vergata''\\
Via della Ricerca Scientifica 1, 00133 Roma, Italy\\
{\em  gianfranco.pradisi@roma2.infn.it}

\end{centering}
\vspace{0.7cm}
$~$\\
\centerline{\bf\Large Abstract}\\

\begin{quote}

The moduli space of toroidal type~I vacua, which are consistent at the non-perturbative level, is composed of independent branches characterized by the number (0, 16 or 32) of rigid branes sitting on top of orientifold planes. This structure persists also when supersymmetry is spontaneously broken {\it \`a la} Scherk--Schwarz. We show that all the components of the moduli space in dimension $D\ge 5$ indeed admit  heterotic dual components, by explicitly constructing  heterotic-type I dual pairs  with the rank of the gauge group  reduced by 0, 8 or 16 units. In the presence of spontaneous breaking of supersymmetry, the dual pairs we consider are also free of tachyonic instabilities at the one-loop level, provided the scale of supersymmetry breaking is lower than the string scale.

\end{quote}

\end{titlepage}
\newpage
\setcounter{footnote}{0}
\renewcommand{\thefootnote}{\arabic{footnote}}
 \setlength{\baselineskip}{.7cm} \setlength{\parskip}{.2cm}

\setcounter{section}{0}


\section{Introduction}

Worldsheet conformal field theories admit marginal deformations. As a consequence, the spectra of string theories possess generically moduli fields at tree level. 
When they are coupled to the visible sector, the arbitrary vacuum expectation values of these massless scalars can spoil the predictability of the models. They also lead to long range forces, which violate the experimental bounds on the validity of the equivalence principle. Therefore, it is important to determine the mechanism(s) responsible for their stabilization and/or for reducing their number from the outset. Among various approaches, string compactifications with fluxes~\cite{fluxes1,fluxes2,fluxes3,fluxes4,fluxes5} or non-perturbative effects at the level of the effective field theories~\cite{gc1,gc2,gc3,gc4,gc5,gc6} have been examined, in order to  lift some of the flat directions of the scalar potential. 

In Refs~\cite{ADLP,Partouche:2019pgv}, an interesting class of models that partially address these issues was analyzed\footnote{In Ref.~\cite{Angelantonj:2006ut} models with spontaneously broken supersymmetry and dynamical stabilization of moduli at self-dual radii in AdS vacua were already considered.}. It was considered in the context of the type I string theory compactified on tori, where supersymmetry is totally but spontaneously broken at a scale $M$, via a stringy version~\cite{openSS1,openSS2,openSS3,openSS4,openSS5,openSS6,openSS7, openSS8,openSS9} of the Scherk--Schwarz mechanism~\cite{SS}. At 1-loop, an effective potential  is generated, and stabilization of all open string Wilson lines was achieved, provided $M$ is lower than the string scale,\footnote{In our conventions, all dimensionfull quantities are expressed in string units, with $\alpha'$ set to 1.}  a fact that we assume throughout the present paper. Indeed, by \mbox{T-dualizing} all the internal directions, one may switch to an orientifold description  where the open string moduli translate into the positions of the 32 Dirichlet-branes (D-branes). The key point is that these objects are either moving in the bulk in pairs as mirror images, or they are rigid, namely their positions are forced to be fixed on top of orientifold planes (O-planes), thus reducing the rank of the gauge group as well as the dimension of the moduli space~\cite{frozbrane}.

However, difficulties persist about the moduli arising from the closed string sector. First of all, these scalars (except $M$ {\it a priori}) remain flat directions of the 1-loop effective potential $\Vone$.  In fact, in $D$ spacetime dimensions, one finds that $\Vone\sim (n_F-n_B) M^D$ at the minima of the open string moduli, where $n_F$ and $n_B$ are  the numbers of massless fermionic and bosonic degrees of freedom, respectively. Second, to account for the flatness of the universe, peculiar models sometimes  referred to as ``super no-scale models'', which  have massless degenerate spectra ($n_F=n_B$), should be considered\cite{nFB0-1,nFB0-2,nFB0-3,nFB0-4,nFB0-5,nFB0-6,nFB0-7}.\footnote{In that case, $\Vone$ is much lower than in generic models, and it is conceivable that it may combine with higher loops corrections to yield a stabilization of $M$ and of the dilaton field, with a small value of the resulting cosmological constant. } Alternatively,  one can deal with Friedmann-Lema\^itre-Robertson-Walker flat cosmological evolutions, where the supersymmetry breaking scale~\cite{CFP,CP,qnsr} and possibly the finite temperature of the Universe~\cite{cosmoT-1,cosmoT-2,cosmoT-3,cosmoT-4,cosmoT-5,cosmoT-6,cosmoT-7} are time-dependent.

In the present work, we reconsider the above class of open string models from a heterotic dual point of view. One of our motivations is that finding from scratch heterotic models free of tachyonic instabilities at 1-loop turns out to be a difficult task, even when $M\ll 1$~\cite{CP}. Hence, an efficient way to reach this goal is to start from the orientifold picture, where models with these properties are easily identified, thanks to the interpretation of moduli in terms of geometrical positions of D-branes. Moreover, since open string models are based on perturbative constructions, additional conditions should exist to  ensure their consistency at the non-perturbative level~\cite{triple,np-2,np-3,np-4,np-5}. In particular, the $O(32)$ gauge bundles should allow an embedding in $Spin(32)/\Z_2$ bundles, as required from the dual (perturbative) heterotic point of view. It turns out that the moduli space of the orientifold models splits into distinct components characterized by various patterns of rigid branes. One of our main results is that at least for $D\ge 5$, all of these consistent branches admit heterotic dual descriptions. Finally, one may forecast that the stabilization of the closed string moduli present in the orientifold picture may be achieved on the heterotic side, at points of enhanced gauge symmetry in moduli space~\cite{GV,cosmoT-4, cosmoT-5, cosmoT-3, cosmoT-6, cosmoT-7}. However, this expectation turns out to be essentially incorrect, due to the fact that the relevant heterotic descriptions are freely acting orbifold constructions that project out the extra states that usually arise at particular points in moduli space. Hence, heterotic and type~I models are typically on equal footing from this point of view. 

The plan of the paper is as follows. In Sect.~\ref{Oset}, we summarize the relevant features of the orientifold models of Ref.~\cite{ADLP}, whose open string moduli are stabilized at 1-loop when $M\ll 1$. We also list the conditions valid in dimension $D\ge 5$~\cite{triple} that are expected to guarantee the non-perturbative consistency of the models. Sect.~\ref{symS} presents the simplest example of a heterotic model that is dual to such an orientifold theory. It is realized in five  dimensions and corresponds, on the open string side, to the case where the 32 D-branes are isolated with rigid positions, thus generating a trivial gauge symmetry we formally denote by  $SO(1)^{32}$. This notation is justified by the fact that in general, a stack of $p$ D-branes on an orientifold plane yields an $SO(p)$ gauge symmetry. On the heterotic side, the counterpart of an isolated rigid D-brane located on top of an orientifold plane is an Ising conformal block. This peculiar model turns out to be free of any tachyonic instability at tree level, as is also the case in the model of Ref.~\cite{Angelantonj:2006ut}, even at large supersymmetry breaking scale. In the $SO(1)^{32}$ case, the potential is positive at least when $M\ll1$, so that $M$ may be attracted to  smaller and smaller values. On the contrary, in Ref.~\cite{Angelantonj:2006ut}, the scale $M$ is stabilized around 1, where it leads to a negative potential at 1-loop. It should be also noticed  that the $SO(1)^{32}$ model, in its supersymmetric version, already appeared in Ref.~\cite{GKP}. In Sect.~\ref{asymS}, we show that in the heterotic description, the existence of points of enhanced gauge symmetry responsible in principle for the stabilization of internal torus moduli is quite limited. Sect.~\ref{so31} is devoted to the construction of another  example of a heterotic model in six dimensions. It is dual to a tachyon free orientifold theory, whose open string gauge symmetry is $[SO(3)\times SO(1)]^8$ coupled to fermionic matter in the ``bifundamental'' representations. The heterotic description  necessarily requires an asymmetric freely acting orbifold projection, as dictated by modular invariance. Our conclusions and perspectives are contained in Sec. \ref{cl}.


\section{Orientifold vacua, rigid branes and broken supersymmetry}
\label{Oset}

Before moving to the construction and analysis of heterotic models with reduced rank, let us review the main features of toroidal orientifold constructions with broken supersymmetry, free of tachyonic instabilities at 1-loop (provided $M\ll1$)~\cite{ADLP}, and that are expected to be consistent at the non-perturbative level~\cite{triple}.  

Our starting point is the type I string in $D$ dimensions, obtained by an orientifold \cite{Sagnotti:1987tw, Bianchi:1990yu, Bianchi:1990tb, Pradisi:1988xd,Angelantonj:2002ct} of the ten-dimensional type II string compactified on a torus $T^{10-D}$.  We denote the Neveu--Schwarz-Neveu--Schwarz (NS-NS) metric and Ramond--Ramond (RR) two-form moduli by $G_{IJ}$ and $C_{IJ}$, $I,J\in\{D,\dots,9\}$. For our purposes, it is convenient to T-dualize all of the internal directions, in order to switch to a perturbatively equivalent orientifold of type IIA when $D$ is odd, or of type IIB when $D$ is even. As known, these orientifolds amount in neutral combinations of O$(D-1)$-planes and D$(D-1)$-branes. Although D-branes are dynamical defects in spacetime which support the open string endpoints, orientifold planes are rigid walls localised at the fixed loci of the involution $\tilde \Omega = \Omega \, I_{10-D}$, where $\Omega$ is the standard worldsheet parity while $I_{10-D}$ inverts the $10-D$ spacelike coordinates transverse to the O$(D-1)$-planes.\footnote{Depending on the value of $D$, an extra operator $(-1)^{F_R}$ is needed in order to ensure that the orientifold involution is indeed order-two. In our notations, $(-1)^{F_R}$ flips the sign of all NS-R and RR states.} The involution $\tilde \Omega$ has $2^{10-D}$ fixed points in the T-dual torus. Each fixed locus supports a different O-plane, whose nature, {\it i.e.} its RR charge and tension, depends on background values of  discrete closed string moduli\footnote{These are the components of the NS-NS metric $\tilde G_{IJ}$ and RR two-form $\tilde C_{IJ}$, $I,J\in\{D,\dots,9\}$,  which are odd with respect to the orientifold involution.} \cite{Bianchi:1991eu,Angelantonj:1999jh,Angelantonj:1999xf,Pradisi:1999ii} which, in turn, are related to the presence or not of vector structure \cite{Witten:1997bs,Bianchi:1997rf}.  Therefore, depending on the number of non-trivial discrete deformations, the orientifold vacuum  involves different numbers of  O$(D-1)_+$- and O$(D-1)_-$-planes with charges and tensions given by $\pm 2^{D-5}$, respectively. A number $N$ of D$(D-1)$-branes is then added to the construction to cancel the total RR charge. The open string moduli are nothing but the positions of the D-branes in the T-dual torus, $\tilde X^I=2\pi a^I_\alpha$, where $\alpha\in\{1, \dots , N\}$ labels the different D-branes. They are dual to the Wilson lines $a_\alpha^I$ of the initial type I theory compactified on $T^{10-D}$. 
Compatibility with the orientifold involution $\tilde\Omega$ implies that the configurations of D-branes are typically given in terms of pairs of branes, with arbitrary coordinates  $\tilde X^I=2\pi a^I_\alpha$ and $\tilde X^I=-2\pi a^I_\alpha$. Alternatively, a brane can sit on top of an orientifold plane, in which case its position is rigid, with $\tilde X^I=2\pi a_\alpha^I\equiv 0$ or~$\pi$.  Indeed, it was argued in \cite{Bianchi:1991eu} that an odd number of D-branes could be moved close to an orientifold plane\footnote{The paper \cite{Bianchi:1991eu} actually discusses the dual construction in terms of generic Wilson lines assumed to be in $O(32)$ rather than $SO(32)$.} yielding, for instance, a Chan-Paton gauge group $SO(15)\times SO(17)$, which is perfectly legitimate from the vantage point of type I perturbation theory. 

As a result, the moduli space admits disconnected components that are characterized by the rank of the open string gauge group, which is lower than 16 when O$(D-1)_+$-planes are present or  D$(D-1)$-branes have rigid locations. In particular, the way rigid branes are distributed among the O-planes does matter when $D\le 7$~\cite{triple}.\footnote{There are topologically inequivalent configurations, {\it i.e.}  not related to each other by a change of coordinates.}  In addition, it is very important to stress that by assuming heterotic-type I duality, non-trivial constraints arise for the non-perturbative consistency of orientifolds \cite{triple,np-2,np-3,np-4,np-5}.  To be specific, let us consider the case where all O$(D-1)$-planes are  O$_-$-planes, so that \mbox{$N=32$}. When rigid D$(D-1)$-branes are present, Wilson line matrices $W_I=\mbox{diag}(e^{2i\pi a_\alpha^I}, \alpha\in\{1,\dots, 32\})$ in the type I picture can have determinant equal to $-1$, and thus correspond to $O(32)$ flat connections on $T^{10-D}$. At the non-perturbative level, the authorized $O(32)$ flat bundles have to lift to $Spin(32)/\Z_2$ bundles, and in the special case of $10-D=3$, the Chern--Simons invariant must also vanish, for the gauge theory to be anomaly free.\footnote{In Ref.~\cite{ADLP}, only the condition that the $O(32)$ bundle is an $SO(32)$ bundle was imposed. } Under these conditions, heterotic dual descriptions are expected to exist. In Ref.~\cite{triple}, all D$(D-1)$-brane configurations satisfying these constraints in dimension  $D\ge 5$ have been classified, provided the O$(D-1)$-planes are all of O$_-$ type, which is the case also considered throughout the present work.  Denoting with $G_{\rm max}$ the gauge group of maximal dimension generated by the open strings and with O$_-^\prime$ the orientifold planes\footnote{We are using here the notation in Ref. \cite{triple}. It should be noticed that the O$_-^\prime$-planes are also known in literature as $\widetilde{\text{O}}_-$-planes. As known, they can be related by chains of $S$ and $T$ dualities to the so called O$_+$-planes and $\widetilde{\text{O}}_+$-planes that, however, are not discussed here.} where rigid branes sit, it turns out that:
\begin{itemize}
\item For $D\ge 7$, rigid branes are forbidden. Therefore, the moduli space has a single branch and $G_{\rm max}=SO(32)$.  

\item For $D=6$, two branches exist. The first one corresponds to the seven-dimensional models compactified on a circle. There are no rigid branes and $G_{\rm max}=SO(32)$. In the second branch, there is a rigid brane sitting on each of the 16 O$5_-^\prime$-planes, so that  $G_{\rm max}=SO(17)\times SO(1)^{15}$, with a reduced rank.

\item For $D=5$, three cases can arise. In the first, no rigid branes are present and $G_{\rm max}=SO(32)$. In the second, exactly 16 rigid branes must sit on top of the 16 O$4_-^\prime$-planes located on one of the hyperplanes $\tilde X^I=0$ or $\pi$, for some $I\in\{5,\dots,9\}$. In this branch, $G_{\rm max}=SO(17)\times SO(1)^{15}$. The previous two branches arise simply by compactifying the $D=6$ allowed models on one additional circle. Finally, a third case exists where all 32 D4-branes are rigid and isolated, each of them being located on one of the 32 O$4_-^\prime$-planes. In this last branch, the open string  ``gauge group''  is $SO(1)^{32}$, which is trivial.   
\end{itemize}

The above observations have interesting consequences when a spontaneous breaking of supersymmetry is implemented at tree-level by the Scherk--Schwarz mechanism~\cite{SS}, generalized to the framework of open string theory~\cite{openSS1,openSS2,openSS3,openSS4,openSS5,openSS6,openSS7}.  In the present paper, we will consider the simplest realization of  such a breaking, whereby  $(-1)^F$ is gauged and fibered along a single compact direction, with $F$ the spacetime fermion number. In the initial type I setup, this amounts to introducing a mass gap $M$ of the order of the inverse length of this  coordinate, between bosonic and fermionic superpartners.
Because no linearly realized supersymmetry is left over, an effective potential depending on all Wilson lines is generated at 1-loop. 
As shown at the perturbative level in Ref.~\cite{ADLP} , the configurations  where all D-branes (rigid or not) in the orientifold  picture are distributed on the O$(D-1)$-planes correspond to extrema of the potential. In particular, let us  denote with $p_{2A-1}$ and $p_{2A}$, $A\in\{1,\dots,2^{10-D}/2\}$, the numbers of D$(D-1)$-branes stacked on the O-planes $2A-1$ and $2A$ that are adjacent along the (T-dualized) Scherk--Schwarz direction at  0 and $\pi$, respectively.
When the supersymmetry breaking scale $M$ is lower than the string scale,  local minima are obtained when each pair $(p_{2A-1},p_{2A})$ takes one of the following values:
\be
(p,0)\, , \;(0,p)\, , \; (p,1)\, , \; (1,p)\, ,  \;\mbox{ except \,$(2,1)$, $(1,2)$}\, ,
\label{cstab}
\ee
with $p$ a positive integer. The Wilson line/position moduli of the stacks of D-branes (not rigid from the outset)  are  then massive at 1-loop, except when $(p_{2A-1},p_{2A})=(2,0)$, $(0,2)$, $(3,1)$ or $(1,3)$, for which  they are flat directions.\footnote{These directions are flat up to exponentially suppressed terms that are no more negligible when the associated pairs of branes move so far that they are approaching other O-planes. When this is the case,  the system is better described  by a new stable or unstable configuration characterized by other $(p_{2A-1},p_{2A})$, $A\in\{1,\dots,2^{10-D}/2\}$.} 

At such local minima, the 1-loop effective potential takes the particularly simple form
\be
\Vone=(n_F-n_B)  \xi_D  M^D+\O\big(M^{D\over 2}e^{- {2\pi\over M}}\big)\;\!, 
\label{rule}
\ee
where $n_F$ and $n_B$ are the number of massless fermionic  and bosonic degrees of freedom. In this expression, $\xi_D>0$ is a constant that captures the contributions of the light towers of Kaluza--Klein (KK) modes propagating along the large Scherk--Schwarz direction (in the type~I picture). Notice that up to the exponentially suppressed terms, when \mbox{$n_F-n_B\neq 0$}, Eq.~(\ref{rule}) depends only on the supersymmetry breaking scale $M$, which is a particular combination of internal metric components.  This means that all other closed string moduli, {\it i.e.} the dilaton, the remaining components of $G_{IJ}$ and the RR two-form moduli $C_{IJ}$, are flat directions. Finally, the numbers of massless states, which are given by 
\be
n_B =8\bigg(8+\sum_{A=1}^{2^{10-D}} \frac{p_{A}(p_{A}-1)}{2}\bigg),\;\;\quad n_F =8\sum_{A=1}^{2^{10-D}/2}p_{2A-1}p_{2A}\,,
\label{nfbtotD'}
\ee
can be derived as follows. The $8\times 8$ bosons originate from the (dimensionally reduced) ten-dimensional dilaton, metric and two-form. The other degrees of freedom counted by $n_B$ correspond to the bosonic parts of vector multiplets in the adjoint representations of the $SO(p_A)$ gauge groups generated by the stacks of $p_A$ D$(D-1)$-branes on the O$(D-1)$-planes. The massless fermions arise from strings  stretched between all pairs  of stacks of D-branes located on adjacent O$(D-1)$-planes along the Scherk--Schwarz direction. Therefore, they are in the bifundamental representation of $SO(p_{2A-1})\times SO(p_{2A})$. 
The reason why they are massless is that the mass arising from the separation along the Scherk--Schwarz direction is exactly compensated by the Scherk--Schwarz mass gap attributed to fermions, as compared to bosons. 

As said, although all sets of  $(p_{2A-1},p_{2A})$ with $A\in\{1,\dots,2^{10-D}/2\}$ satisfying the conditions given in~\eqref{cstab} yield  perturbatively allowed local minima of the 1-loop potential with respect to the open string Wilson lines, the number of choices consistent at the non-perturbative level is more restricted.  For instance, following what stated previously, in dimensions $D\ge 7$ only the solutions $(p,0)$ and $(0,p)$ with even $p$'s  should be authorized, corresponding to the single allowed branch of the moduli space. On the other hand,  in $D=6$ and $D=5$ more choices are expected to be non-perturbatively valid, reflecting the existence of the two or three branches of the moduli space. In the following, we shall show that this is indeed the case by constructing explicit heterotic backgrounds with broken supersymmetry that are dual to tachyon free orientifold configurations consistent non-perturbatively.


\section{Heterotic \bm $SO(1)^{32}$ model in five dimensions}
\label{symS}

 In this section, we consider the heterotic model that is probably the simplest one providing a dual description of an  orientifold theory with rigid branes. It illustrates the use of Ising conformal blocks in the derivation of modular invariant partitions functions, and was first described in its supersymmetric version in Ref.~\cite{GKP}. In the class of open string theories considered in Ref.~\cite{ADLP}, the model we focus on corresponds to the case where all 32 D4-branes are rigid and separated on the 32 O$4_-^\prime$-planes, with $SO(1)^{32}$ open string ``gauge group''. Its moduli space is nothing but one of the three branches allowed in five dimension, as described in the previous section. 
The model is also characterized by the greatest possible value of $n_F-n_B$, which is positive.  The heterotic dual description is (relatively) simple in the sense that it can be realized in terms of a freely acting orbifold that is  left-right symmetric,  {\it i.e.} geometric.

Let us consider a five-dimensional heterotic model based on a $\Z_2^5$ free orbifold action on the internal $T^5$ that also breaks completely the $SO(32)$ gauge symmetry and supersymmetry. To be specific, each orbifold generator $G_I$, $I\in\{5,\dots,9\}$, acts as follows:
\begin{itemize}
\item It implements a half-period shift of the compact direction  $X^I\equiv X_L^I+X_R^I$, where $X^I_L$, $X^I_R$ are the left- and right-moving pieces.

\item $G_9$ also contains an action $(-1)^F$, whose effect is to implement the Scherk--Schwarz spontaneous breaking of supersymmmetry~\cite{SSstring-1,SSstring-2,SSstring-3,SSstring-4,SSstring-5}.

\item The $G_I$'s twist the extra 32 real fermions of the right-moving bosonic side of the heterotic string in such a way that all of them have distinct boundary conditions. The actions of the 5 generators are shown in Table~\ref{T1}. They imply the initial conformal block generating the $SO(32)$ degrees of freedom to be replaced by 32 copies of combinations of Ising characters.

\end{itemize}
\begin{table}[h]
\begin{center}
\begin{tabular}{|c|l|}
\hline
$G_9$ & \px\px\px\px\px\px\px\px\px\px\px\px\px\px\px\px\x\x\x\x\x\x\x\x\x\x\x\x\x\x\x\x\\
$G_8$ & \px\px\px\px\px\px\px\px\x\x\x\x\x\x\x\x 	\px\px\px\px\px\px\px\px\x\x\x\x\x\x\x\x \\
$G_7$ & \px\px\px\px\x\x\x\x \px\px\px\px\x\x\x\x \px\px\px\px\x\x\x\x \px\px\px\px\x\x\x\x 	\\
$G_6$ & \px\px\x\x \px\px\x\x \px\px\x\x \px\px\x\x \px\px\x\x \px\px\x\x \px\px\x\x \px\px\x\x 	\\
$G_5$ & \px \x \px \x \px \x \px \x \px \x \px \x \px \x \px \x \px \x \px \x \px \x \px \x \px \x \px \x \px \x \px \x \\
\hline
\end{tabular}
\caption{\label{T1} \em \footnotesize  Twist actions of the five generators $G_I$ on the 32 right-moving real worldsheet fermions. A ``$-$'' sign indicates a non-trivial $\Z_2$ twist of the corresponding fermion.}
\end{center}
\vspace{-0.5cm}
\end{table}

What we are interested in is the 1-loop effective potential 
\be
\Vone=-{1\over (2\pi)^5}\int_{\F} {\d^2 \tau\over 2\tau_2^2} \, Z_{(5)}\, , 
\ee
where $\F$ denotes the $SL(2,\Z)$ fundamental domain, $\tau=\tau_1+i\tau_2$ is the genus-1 Techm\"uller parameter, and $Z_{(5)}$ is the partition function. In the present case, we have
\be
\begin{aligned}
Z_{(5)}&={1\over \big(\sqrt{\tau_2}\eta\bar \eta\big)^3}\, {1\over 2^5}\sum_{\vec h,\vec g}{\Gamma_{5,5}\!\left[{}^{\vec h}_{\vec g}\right]\over (\eta\bar \eta)^5}\, {1\over 2}\sum_{a,b}(-1)^{2(a+b+2ab)}{\theta[{}^a_b]^4\over \eta^4}\, {\Gamma_{0,16}\!\left[{}^{\vec h}_{\vec g}\right]\over \bar \eta^{16}}\, (-1)^{4(g_9a+h_9b+g_9h_9)}\\
&\equiv {1\over 2^5}\sum_{\vec h,\vec g}Z_{(5)}\!\left[{}^{\vec h}_{\vec g}\right], 
\end{aligned}
\label{Z}
\ee
where $a,b$ and the components of the 5-vectors $\vec h,\vec g$ take the values $0$ or ${1\over 2}$. In our notations, $\eta$ and $\theta$ denote the Dedekind and Jacobi modular functions, $a,b$ are the spin structures of the left-moving worldsheet fermions (where $2a\equiv F$), $\vec h$ labels the 32 (un)twisted sectors, while the sums over $b$ and $\vec g$ implement the GSO and orbifold projections. Because the $\Z_2^5$ generators are freely acting, all 31 twisted sectors are massive. Therefore, the gauge symmetry generated by the 32 right-moving real fermions arises solely from the untwisted sector, and it realizes the following chain of breakings 
\be
SO(32)\overset{G_9}{\longrightarrow} SO(16)^2\overset{G_8}{\longrightarrow} SO(8)^4\overset{G_7}{\longrightarrow} SO(4)^8\overset{G_6}{\longrightarrow} SO(2)^{16}\overset{G_5}{\longrightarrow}SO(1)^{32}\, ,
\ee 
where $SO(1)$ denotes the trivial group containing only the neutral element. 
As a result, no marginal deformation (no Wilson line) is allowed by the shifted $\Gamma_{0,16}\big[{}^{\vec h}_{\vec g}\big]$ Narain lattices and 
this setup is expected to be dual to the orientifold  configuration with 32 rigid branes.  To be specific, from the twist actions of Table~\ref{T1}, we have
\be
{\Gamma_{0,16}\!\left[{}^{\vec h}_{\vec g}\right]\over \bar \eta^{16}}= {1\over 2}\sum_{\gamma,\delta}{\bar\theta[{}^\gamma_\delta]^{1\over 2}\over \bar\eta^{1\over 2}}\, {\bar\theta\big[{}^{\gamma+h_5}_{\delta+g_5}\big]^{1\over 2}\over \bar\eta^{1\over 2}}\, \cdots {\bar\theta\big[{}^{\gamma+h_5+\cdots+h_9}_{\delta+g_5+\cdots+g_9}\big]^{1\over 2}\over \bar\eta^{1\over 2}}\, ,
\label{16}
\ee
where the spin structures $\gamma,\delta$ take values $0,{1\over 2}$, and each of the 32 factors is a single Ising conformal block.  On the contrary, the lattice of zero-modes of the internal torus depends on the metric and antisymmetric tensor moduli $G_{IJ}$ and $B_{IJ}$,
 \be
 \begin{aligned}
\Gamma_{5,5}\!\left[{}^{\vec h}_{\vec g}\right]\!&={\sqrt{\det G}\over \tau_2^{5\over 2}}\sum_{\vec \ell,\vec n} e^{-{\pi\over \tau_2}[\ell_I+g_I+(n_I+h_I)\bar \tau](G+B)_{IJ}[\ell_J+g_J+(n_J+h_J)\tau]}\\
&= \sum_{\vec m,\vec n}e^{2i\pi \vec g\cdot\vec m}\, \Lambda_{\vec m, \vec n+\vec h} \, ,
 \end{aligned}
\label{ga}
 \ee
where $\vec m,\vec n\in \Z^5$ are the momenta and winding numbers, $\vec \ell\in \Z^5$, and  
\begin{equation}
\Lambda_{\vec m, \vec n}= q^{{1\over 4}P^L_IG^{IJ}P^L_J}\, \bar q^{{1\over 4}P^R_IG^{IJ}P^R_J}\, ,
\label{lattice1}\end{equation}
where we have defined 
\begin{equation}
P^L_I=m_I+(B+G)_{IJ}\, n_J\, , \quad  P^R_I=m_I+(B-G)_{IJ}\, n_J\, ,\quad q=e^{2i\pi \tau}\, .
\label{lattice2}
\end{equation}
In Eq.~(\ref{Z}), the sign $(-1)^{4(g_9a+h_9b+g_9h_9)}$ is responsible for the spontaneous breaking of supersymmetry~\cite{SSstring-5}. Notice that it reverses the GSO projection in the 16 twisted sectors that  have $h_9={1\over 2}$. 

In the untwisted sector $\vec h=\vec 0$, we obtain 
\be
\begin{aligned}
Z_{(5)}\!\left[{}^{\vec 0}_{\vec g}\right]\!={1\over \tau_2^{3\over 2}(\eta\bar\eta)^8}\,& \Big(V_8-(-1)^{g_9}S_8\Big) \sum_{\vec m,\vec n}(-1)^{2\vec g\cdot \vec m}\Lambda_{\vec m,\vec n}\\
&\Big( \bar O_{16}\bar O_{16}-(-1)^{\delta_{\vec g,\vec 0}}\bar V_{16}\bar V_{16}+\bar S_{16}\bar S_{16}-(-1)^{\delta_{\vec g,\vec 0}}\bar C_{16}\bar C_{16}\Big),
\end{aligned}
\label{Zhg}
\ee
where $O(2n)$ affine characters are defined as~\cite{Bianchi:1990yu,Angelantonj:2002ct} 
\begin{align}
O_{2n}&={\theta\big[{}^0_0\big]^n+\theta\Big[{}^{{}^{\scriptstyle \phantom{.}\!0}}_{1\over 2}\Big]^n\over 2\eta^n}\, , &V_{2n}&={\theta\big[{}^0_0\big]^n-\theta\Big[{}^{{}^{\scriptstyle \phantom{.}\!0}}_{1\over 2}\Big]^n\over 2\eta^n}\, ,\nonumber \espD\\
S_{2n}&={\theta\Big[{}^{1\over 2}_{\phantom{\!|}_{\scriptstyle0}}\Big]^n+(-i)^n\theta\Big[{}^{1\over 2}_{1\over 2}\Big]^n\over 2\eta^n}\, ,
&C_{2n}&={\theta\Big[{}^{1\over 2}_{\phantom{\!|}_{\scriptstyle0}}\Big]^n-(-i)^n\theta\Big[{}^{1\over 2}_{1\over 2}\Big]^n\over 2\eta^n}\, .
\label{charac}
\end{align} 
The $O(16)$  characters arise because the 31 orbifold group elements $\prod_I G_I^{2g_I}$, for $\vec g\neq \vec 0$,  twist 16 out of the 32 real right-moving fermions, as can be checked from Table~\ref{T1}. Therefore, the  characters with  different $\vec g\neq \vec 0$ are formally equal. As announced before, the shift actions of the generators $G_I$ on the periodic directions $X^I$, $I\in\{5,\dots,9\}$, imply  that all states in the twisted sectors $\vec h\neq \vec 0$ have very large masses, when the sizes of the compact directions $X^5, \dots, X^8$ are not very small with respect to the string length. To be specific, we have 
\be
\begin{aligned}
Z_{(5)}\!\left[{}^{\vec h}_{\vec 0}\right]\!={1\over \tau_2^{3\over 2}(\eta\bar\eta)^8}\,& \!\Big[\delta_{h_9,0}\big(V_8-S_8\big) +\delta_{h_9,{1\over 2}}\big(O_8-C_8\big)\Big]\sum_{\vec m,\vec n}\Lambda_{\vec m,\vec n+\vec h}\\
&\Big( \bar O_{16}\bar S_{16}+\bar V_{16}\bar C_{16}+\bar S_{16}\bar O_{16}+\bar C_{16}\bar V_{16}\Big),
\end{aligned}
\label{Ztw}
\ee
where $\vec n+\vec h$ cannot vanish. In this expression, the $O(16)$ characters are implicitly dependent on $\vec h\neq \vec 0$. 
Notice that the right-moving combinations $\bar O_{16}\bar S_{16}/\bar \eta^8$, $\bar S_{16}\bar O_{16}/\bar \eta^8$ start at the massless level, implying the absence of tachyonic modes at any point in the moduli space. This is to be contrasted with the case of non-freely acting orbifold actions. 
To clarify this point, consider for instance a single twist action $G_I$ on the $\Gamma_{0,16}$ lattice. This orbifold action  breaks $SO(32)\to SO(16)\times SO(16)$ in the untwisted sector $h_I=0$. If $G_I$ was not acting as a shift along $X^I$, there would be additional massless states arising in the twisted sector $h_I={1\over 2}$, in spinorial representations of the $SO(16)$'s. This mechanism is well known, since it induces the restauration of an $E_8\times E_8$ gauge symmetry \cite{DHVW2} due to the identity
\be
\Big(\bar O_{16}+\bar S_{16}\Big)\times  \Big(\bar O_{16}+\bar S_{16}\Big)= \bar O_{E_8}\times \bar O_{E_8}\, , 
\ee
where $O_{E_8}$ is the affine $E_8$ character. The ground state of the $E_8\times E_8$ Kac-Moody algebra turns out to have negative conformal weight, $\bar O_{16}\bar O_{16}/\bar \eta^8=\bar q^{-1}+\cdots$. Combined with  the NS left-handed ground state, $O_8/\eta^8=q^{-{1\over 2}}+\cdots$,  level-matched tachyons arise in regions of moduli space where the supersymmetry breaking scale $M$, which is defined as  
\be
M={\sqrt{G^{99}}\over 2} \, ,
\ee
is of the order of the string scale. 

The light spectrum of the model when $M\ll 1$ amounts to the massless states accompanied with their KK towers of modes propagating along the large Scherk--Schwarz direction $X^9$. At generic points of the $T^5$ moduli space,\footnote{Actually, we will see shortly that there are no extra massless states at any particular point of the moduli space.} the massless bosonic degrees of freedom arise from the combinations of characters $V_8\bar O_{16}\bar O_{16}/(\eta\bar\eta)^8$ and $V_8 \bar V_{16}\bar V_{16}/(\eta\bar\eta)^8$, in the untwisted sector $\vec h=\vec 0$. Their counting goes as follows,
\be
\begin{aligned}
n_B&={8\over 2^5}\Big( \big[8+120+120+16\times 16\big]+31 \times \big[8+120+120-16\times 16\big]\Big)\\
&= 8\times 8\, ,
\end{aligned}
\ee
which matches with the orientifold result given in Eq.~(\ref{nfbtotD'}), when $p_A=1$, $A\in\{1,\dots,32\}$. These states correspond to the bosonic parts of the $\N_5=2$ supergravity multiplet and of five Abelian vectors multiplets, in five dimensions. In total, a gauge symmetry $U(1)^5_{\rm grav}\times U(1)^5$ is generated by  $(G+B)_{I\mu}$ and $(G+B)_{\mu I}$, $I\in\{5,\dots,9\}$, where the first factor is associated with the graviphotons. The massless fermions arise from the characters $-S_8\bar O_{16}\bar O_{16}/(\eta\bar\eta)^8$ and $-S_8 \bar V_{16}\bar V_{16}/(\eta\bar\eta)^8$. Since the generator $G_9$ contains an action $(-1)^F$, the sectors with $g_9={1\over 2}$ contribute with an opposite sign, giving
\be
\begin{aligned}
n_F&={8\over 2^5}\Big( \big[8+120+120+16\times 16\big]+(15-16) \times \big[8+120+120-16\times 16\big]\Big)\\
& = 8\times 16\, .
\end{aligned}
\ee
All of these states are neutral with respect to the gauge group $U(1)^5_{\rm grav}\times U(1)^5$. As before, the value of $n_F$ agrees with Eq.~(\ref{nfbtotD'}), when all $p_A$'s are equal to 1. This confirms that 
the heterotic model is dual to the orientifold theory with 32 rigid D4-branes   on top of the 32 O4-planes. 

Some comments related to the moduli fields arising in our dual pair are in order.  In the initial type I framework, the moduli are the dilaton and the internal metric $G_{IJ}$ in the NS-NS sector, and the internal RR 2-form $C_{IJ}$. As can be seen in Eq.~(\ref{rule}), it turns out that all of them (except $M$) remain massless at 1-loop, up to  contributions suppressed exponentially when $M\ll 1$. To understand why, notice that $G_{IJ}$ and $C_{IJ}$ can be interpreted as Wilson lines along $T^5$ of the Abelian vector bosons $G_{\mu J}$ and $C_{\mu J}$ present in ten dimensions. Because the states lighter than $M$ in the open string perturbative spectrum are neutral with respect to these gauge bosons, their masses at tree level are independent on the Wilson lines $G_{IJ}$, $C_{IJ}$. Hence, the 1-loop Coleman-Weinberg effective potential, which is exclusively expressed in terms of the classical squared masses, admits flat directions associated with these scalars, up to exponentially suppressed corrections arising from the heavy spectrum. 

On the contrary, one may a priori  expect that the dual moduli fields $G_{IJ}$, $B_{IJ}$ be stabilized in the heterotic model,  because of the existence of perturbative states charged under $G_{\mu I}$, $B_{\mu I}$. Actually, such charged states must have  non-trivial winding numbers along $T^5$, and thus correspond to non-perturbative D1-branes in type I string, where they cannot lead to a perturbative stabilization of moduli fields. On the heterotic side, one can show that the Coleman-Weinberg effective potential is extremal with respect to a Wilson line,  when the latter takes a value at which non-Cartan states charged under the associated Abelian symmetry are becoming massless.\footnote{This follows from the fact that at such points, the tadpoles of the Wilson lines are proportional to the sum over the charges, and that all particles can be paired with their antiparticles to yield vanishing contributions~\cite{GV}.} This mechanism was used in Refs~\cite{GV, cosmoT-4, cosmoT-5, CP} to stabilize internal radii in toroidal compactifications. Therefore, in our case of interest, stabilization of $G_{IJ}$, $B_{IJ}$ relies on the existence of points in the freely acting orbifold  moduli space, where $U(1)^5$ is enhanced to a non-Abelian gauge group.  

To figure out if this is possible, let us first consider the case where the $\Z_2^5$ orbifold action is not implemented.  The states at lowest oscillator levels in the NS sector and with momentum and winding numbers $m_5=-n_5=\pm1$ (and $m_I=n_I=0$, $I\neq 5$) are the bosonic degrees of freedom of two vector multiplets 
of charges $\pm \sqrt{2}$ under the $U(1)$ gauge symmetry associated with the internal direction~5. At the locus in moduli space where $G_{5I}=\delta_{5I}$, $B_{5I}=0$, $I\in\{5,\dots,9\}$, they are becoming massless and enhance $U(1)\to SU(2)$. However, in our model, the generator $G_5$ acts as a half-period shift along $X^5$, implying a projection of the spectrum onto modes with even momenta $m_5$. In particular, the above mentioned non-Cartan massless states are projected out, as follows from Eq.~(\ref{Zhg}),  
\be
{1\over 2}\sum_{g_5=0,{1\over 2}}(-1)^{2g_5m_5}|m_5=-n_5=\pm 1\rangle =0\, ,
\label{proj}
\ee
jeopardizing  the enhancement.  Note that naively, because the generator $G_5$ implements a half-period shift along $X^5$, one may think that an enhanced $SU(2)$ symmetry should arise at the point $G_{5I}=4\delta_{5I}$,  $B_{5I}=0$ in moduli space. However, if this were true, the additional massless states would arise from the twisted sector $h_5={1\over 2}$, and we have seen  in Eq.~(\ref{Ztw}) that this sector is massive, due to the simultaneous action of the generator $G_5$ on the $\Gamma_{0,16}$ lattice. 

The projection of the non-Cartan massless states is not specific to the $SU(2)$ case. 
 Therefore, all internal moduli $G_{IJ}$, $B_{IJ}$ of the heterotic model remain  massless at 1-loop, because perturbative states charged under the Abelian $U(1)^5$ gauge symmetry are not present. In order to stabilize some of the compactification moduli, we have to implement alternative free actions of the $\Z^5_2$ generators on the internal directions, so that non-Cartan states belonging to the untwisted sector $\vec h=\vec 0$ survive. This is what we do in the following section. 
 

\section{Partial heterotic stabilization of the torus moduli}
\label{asymS}

As explained at the end of the previous section,  in order to stabilize some components of the internal metric $G_{IJ}$ and of the antisymmetric tensor $B_{IJ}$ in the heterotic model, we need to imagine different free actions of the generators $G_I$, $I\in\{5,\dots, 9\}$, on the internal torus.  The new orbifold projection should  allow the enhanced gauge symmetry states to survive in the untwisted sector $\vec h=\vec 0$. 

To understand how this can be achieved, it should be noticed that in Eq.~(\ref{proj}), the non-Cartan states of $SU(2)$ would be preserved by the action of the generator $G_5$ if $m_5$ in the exponent was replaced by $m_5+n_5$, which is even. Hence, free actions implemented as half-period shifts on both coordinates $X_L^I+X_R^I$ and T-dual coordinates $X_L^I-X_R^I$ may lead to a stabilization of internal moduli at points of enhanced gauge symmetry. At such loci, $n_F-n_B$ decreases, implying the heterotic massless spectrum to differ from that of the dual perturbative one in type~I.

In orbifold language, actions of this kind on coordinates and T-dual coordinates are said left-right asymmetric~\cite{asymmorb}.  The construction of orbifold models in this case turns out to be more constrained than in the symmetric one, due to modular invariance \cite{modasym-1,modasym-2}. To understand why, let us consider a $(d,d)$-lattice associated with arbitrary shift actions on the  $X_L^I+X_R^I$ and  $X_L^I-X_R^I$, $I\in\{10-d,\dots,9\}$. Labelling the components of the lattice with $d$-vectors $\vec \goh, \vec \gog$ and $\vec \goh',\vec \gog'$, and parameterizing the continuous deformations by a metric $G_{IJ}$ and an antisymmetric tensor $B_{IJ}$, we have  
\be
\begin{aligned}
\Gamma_{d,d}\!\left[{}^{\vec \goh,\vec \goh'}_{\vec \gog,\vec \gog'}\!\right]\!&={\sqrt{\det G}\over \tau_2^{d\over2}}\sum_{\vec \ell, \vec n} e^{2i\pi (\vec \gog'\cdot \vec n-\vec \goh'\cdot \vec \ell)}\, e^{-{\pi\over \tau_2}[\ell_I+\gog_I+(n_I+\goh_I)\bar \tau](G+B)_{IJ}[\ell_J+\gog_J+(n_J+\goh_J)\tau]}\\
&= e^{2i\pi \vec \gog\cdot\vec \goh'}\sum_{\vec m,\vec n}e^{2i\pi (\vec \gog\cdot \vec m+\vec \gog'\vec n)}\, \Lambda_{\vec m+\vec \goh',\vec n+\vec \goh} \, ,
\end{aligned}
\label{pg}
\ee
where the second equality is obtained by Poisson summation over $\vec \ell\in\Z^d$. 
In our case of interest, all components of $\vec \goh, \vec \gog$ and $\vec \goh', \vec \gog'$ take values in ${1\over 2}\Z$.\footnote{The identity in Eq.~(\ref{pg}) is actually valid for arbitrary real vectors $\vec \goh, \vec \gog$ and $\vec \goh', \vec \gog'$.} In the above formula, we make use of the definitions given in Eqs. (\ref{lattice1}), (\ref{lattice2}), but for a $d$-dimensional  torus. Notice that the components of $\vec \goh', \vec \gog'$ can actually be defined modulo 1. However, the most general transformation $\vec \goh\to \vec \goh+\vec \delta$ (or $\vec \gog\to\vec  \gog+\vec \delta$), where $\delta_I\in\{0,1\}$, amounts to multiplying the lattice by $(-1)^{2\vec \gog'\cdot\vec \delta}$ (or $(-1)^{2\vec \goh'\cdot\vec \delta}$). Therefore, in order to construct modular invariant partition functions, the allowed vectors $\vec \goh,\vec \gog,\vec \goh',\vec \gog'$ (and correspondingly the allowed vectors $\vec \delta$) should be constrained for all signs $(-1)^{2\vec \gog'\cdot\vec \delta}$ and $(-1)^{2\vec \goh'\cdot\vec \delta}$ to be $+1$. 

Besides the case of pure left-right symmetric (momentum) shifts we considered in Sect.~\ref{symS}, where all asymmetric vectors vanish, $\vec \goh'= \vec \gog'=\vec 0$, a non-trivial solution is to choose the generator $G_5$ to act on the momenta and  winding numbers of the pair of coordinates $X^4$ and $X^5$ of the  $(6,6)$ lattice of a six-dimensional internal torus. Moreover, the remaining generators $G_6,\dots ,G_9$ can be chosen to be identical to those introduced in the previous section. Namely, $G_6,\dots, G_8$ act on the momenta of $X^6,\dots,X^8$, while $G_9$ implements the Scherk--Schwarz breaking of supersymmetry by acting both on the momentum of the direction $X^9$ and as $(-1)^F$~\cite{SSstring-5}.  Hence, the spacetime dimension of the model is four. The components of the constrained left-right symmetric and left-right asymmetric vectors are 
\be
\begin{aligned}
\label{shi}
&\goh_4=\goh'_4=h_5\, , && \gog_4= \gog'_4=g_5\, ,\\
&\goh_5=\goh'_5=h_5\, , && \gog_5= \gog'_5=g_5\, ,\\
&\goh_I=h_{I}\, , \quad  \goh'_I=0\, , &\quad & \gog_I=g_{I}\, ,\quad \gog'_I=0 \, , \quad I\in\{6,7,8,9\}\, , 
\end{aligned}
\ee
and the shifted lattice can be written as 
 \be
 \begin{aligned}
\Gamma_{6,6}\!\left[{}^{\vec h}_{\vec g}\right]\!={\sqrt{\det G}\over \tau_2^3}\sum_{\vec \ell,\vec n} & (-1)^{2g_5 (n_4+n_5)+2h_5 (\ell_4+\ell_5)}\\
&\times e^{-{\pi\over \tau_2}[\ell_I+\gog_I+(n_I+\goh_I)\bar \tau](G+B)_{IJ}[\ell_J+\gog_J+(n_J+\goh_J)\tau]}\, ,
 \end{aligned}
\label{ga'}
 \ee
where $\vec \ell$, $\vec n$ are 6-vectors. We stress that the 6-vectors $\vec \goh,\vec \gog$ depend only on 5-vectors $\vec h,\vec g$. The key point is that because the transformation  $h_5\to h_5+ 1$ (or $g_5\to g_5+1$) shifts pairs of integers $n_4,n_5$  (or $\ell_4,\ell_5$), they are symmetries of the lattice, as required to construct modular invariant partition functions based on a finite number (equal to $2^5\times 2^5$) of conformal blocks $(\vec h,\vec g)$. In fact, the definitions~(\ref{shi}) mean that each $G_I$, $I\in\{6,\dots,9\}$, acts as a half-period shift of the coordinate $X_L^I+X_R^I$, while $G_5$ acts as half-period shifts on $X_L^4+X_R^4$, $X_L^5+X_R^5$, as well as on the T-dual coordinates $X_L^4-X_R^4$, $X_L^5-X_R^5$.
The Hamiltonian form of the lattice is obtained from Eq.~(\ref{pg}) and results in 
\be
\Gamma_{6,6}\!\left[{}^{\vec h}_{\vec g}\right]\!= (-1)^{4g_5h_5}\, \sum_{\vec m,\vec n}(-1)^{2g_5(m_4+n_4+m_5+n_5)+2\sum_{I=6}^9g_Im_I}\, \Lambda_{\vec m+\vec  \goh', \vec n+\vec \goh} \, .
\label{ga''}
 \ee
Recalling that the generators $G_I$, $I\in\{5,\dots,9\}$, also act as twists on the right-moving worldsheet fermions, the modular invariant partition function can be written as 
\be
\begin{aligned}
Z_{(4)}&={1\over \big(\sqrt{\tau_2}\eta\bar \eta\big)^2}\, {1\over 2^5}\sum_{\vec h,\vec g}{\Gamma_{6,6}\!\left[{}^{\vec h}_{\vec g}\right]\over (\eta\bar \eta)^6}\, {1\over 2}\sum_{a,b}(-1)^{2(a+b+2ab)}{\theta[{}^a_b]^4\over \eta^4}\, {\Gamma_{0,16}\!\left[{}^{\vec h}_{\vec g}\right]\over \bar \eta^{16}}\, (-1)^{4(g_9a+h_9b+g_9h_9)}\\
&\equiv {1\over 2^5}\sum_{\vec h,\vec g}Z_{(4)}\!\left[{}^{\vec h}_{\vec g}\right] , 
\end{aligned}
\label{Z4}
\ee
where the $(0,16)$-lattice is given in Eq.~(\ref{16}).

In the untwisted sector $\vec h=\vec 0$, the NS states at lowest oscillator levels  and with quantum numbers 
\be
\begin{aligned}
& m_4=-n_4=\pm1\, , \quad m_I=n_I=0\, , \quad I\in\{5,6,7,8,9\}\, , \\
\mbox{and}\;\; \quad &m_5=-n_5=\pm1\, , \quad m_I=n_I=0\, ,\quad I\in\{4,6,7,8,9\}\, , \\
\end{aligned}
\ee
 survive the orbifold $\Z_2^5$ projections, since 
 \be
{1\over 2^5}\sum_{\vec g}(-1)^{2g_5(m_4+n_4+m_5+n_5)+2\sum_{I=6}^9g_Im_I}|m_I=-n_I=\pm 1\rangle =|m_I=-n_I=\pm 1\rangle\,  ,\;\;\;  I\in\{4,5\} .
\label{proj2}
\ee
As a result, when the background $(G+B)_{IJ}$ is of the form 
\be
\begin{aligned}
&(G^{(0)}+B^{(0)})_{\alpha\beta}=\left(\!\!\begin{array}{cc}1 & 0\\0&1\end{array}\!\!\right)\! , && \alpha,\beta\in\{4,5\}\, ,\\
&(G^{(0)}+B^{(0)})_{\alpha B}=(G^{(0)}+B^{(0)})_{A\beta}=0\, ,  &&A,B\in\{6,\dots,9\}\, , \\
&(G^{(0)}+B^{(0)})_{AB}\quad \mbox{arbitrary}\, , 
\end{aligned}
\label{backg}
\ee
the bosonic states of $2+2$ vector multiplets are becoming massless, extending the $U(1)^2$ gauge symmetry associated with the directions $X^4$ and $X^5$ to $SU(2)^2$.  

To show that internal moduli fields are stabilized at this locus, we consider the 1-loop effective potential  
\be
\Vone=-{1\over (2\pi)^4}\int_{\F} {\d^2 \tau\over 2\tau_2^2} \, Z_{(4)}\, .
\ee
At low supersymmetry breaking scale compared to the string scale, the dominant contribution of $\Vone$ arises from the $n_F+n_B+\Delta n_B$ massless states, their superpartners, and their KK towers of modes propagating along the large Scherk--Schwarz direction $X^9$,  where $\Delta n_B=8\times (2+2)$ denotes the number of extra massless bosons. All other string states, whose masses are of the order of the string scale, yield exponentially suppressed contribution, $\O(e^{-\pi/M})$.   In Ref.~\cite{CP},  the 1-loop effective potential in the case of a pure toroidal compactifications (no orbifold action, free or not free) is derived. Because we have seen in Sect.~\ref{symS} that all light states of the  $\Z_2^5$ freely acting orbifold model we consider belong to the untwisted sector, $\vec h=\vec 0$, it turns out  that the effective potential in this case equals the result of Ref.~\cite{CP}, up to exponentially suppressed corrections.\footnote{The contribution of the light KK towers of states in the ``non-orbifolded'' model must be dressed by an overall factor $1/2^5$, which is compensated by the sum over $\vec g$. This follows from the fact that none of these towers is projected out. } In order to write the final answer, we define small background fluctuations $y_{IJ}$, $I,J\in\{4,\dots,9\}$, $(I,J)\neq (9,9)$, 
\be
(G+B)_{IJ}=\!\left(\!\!\begin{array}{lc} (G^{(0)}+B^{(0)})_{ij}+\sqrt{2}\, y_{ij}& (G^{(0)}+B^{(0)})_{i9}+\sqrt{2}\, y_{i9}\\(G^{(0)}+B^{(0)})_{9j}+\sqrt{2} \, y_{9j}&(G+B)_{99}\end{array}\!\!\right)\! ,\;\; i,j\in\{4,\dots,8\}\, , 
\ee
in terms of which the Taylor expansion of the 1-loop effective potential at quadratic order and in the string frame takes the form 
\be
\begin{aligned}
\Vone= &\, \big[n_F-(n_B+\Delta n_B)\big]  \xi_{4}\,  M^4 \\
&\, +  M^4 \, {2 \over \pi}\,\xi_{2}\sum_{j=4}^5 8\, T_{[3]_{SU(2)}}\Big[3y_{9j}^2+{1\over G^{99}}\sum_{i=4}^8y_{ij}^2+\cdots \Big]  +\O\big(M^{2}e^{-{\pi/M}}\big).
\label{final}
\end{aligned}
\ee
In this expression, $\xi_D$ is given by 
\be
\xi_{D}={2\, \Gamma({D+1\over 2})\, \zeta(D+1)\over \pi^{3D+1\over 2}}\, \Big(1-{1\over 2^{D+1}}\Big), 
\ee
while the factor 8 counts the degeneracy of the bosons (or fermions) in each vector multiplet, and $T_{[3]_{SU(2)}}=2$ is the Dynkin index of the adjoint representation of $SU(2)$.
In fact, $y_{IJ}$ can be interpreted as the Wilson line along the periodic direction $X^I$, of the Cartan $U(1)$ associated with the direction $X^J$. We see that all Wilson lines along $T^6$ of the enhanced gauged group $SU(2)^2$  acquire a mass at 1-loop, namely the moduli $y_{I4}, y_{I5}$, $I\in\{4,\dots,9\}$, while all others remain massless. Because $n_F-(n_B+\Delta n_B)=8\times (16-8-2-2)=8\times 4$, the potential is positive and the supersymmetry breaking scale $M=\sqrt{G^{99}}/2$  has a tadpole. 

Notice that the preservation of enhanced gauge symmetry states by a generator acting on both momenta and winding numbers is not automatic. As an example, let us consider before implementation of the $\Z_2^5$ orbifold projection  the NS states at lowest oscillator levels that have quantum numbers 
\be
\begin{aligned}
& (m_4,n_4,m_5,n_5)=(\pm1,0,\pm1, \mp 1)\, , \quad (0,\pm1,\pm1, \mp 1)\, , \quad (\pm1,\mp1,0, 0)\, , \\
&\with \quad m_I=n_I=0\, , \quad I\in\{6,7,8,9\}\, .
\end{aligned}
\label{sp3}
\ee
All of these modes turn out to become massless at the point in moduli space corresponding to the background similar to that given in Eq.~(\ref{backg}), but with 
\be
(G^{(0)}+B^{(0)})_{\alpha\beta}=\left(\!\!\begin{array}{cc}1 & 1\\0&1\end{array}\!\!\right)\! , \qquad  \alpha,\beta\in\{4,5\}\, .
\ee
In that case, the $U(1)^2$ gauge symmetry associated with the directions $X^4$ and $X^5$ is enhanced to $SU(3)$.
However, among these six non-Cartan states of $SU(3)$, the four first are projected out by the operator appearing in the l.h.s. of Eq.~(\ref{proj2}). Therefore, the asymmetric orbifold action of the generator $G_5$ reduces $SU(3)$ to $SU(2)$, and we obtain nothing better than what we would have found, had we compactified the model of Sect.~\ref{symS} on a factorized circle of radius 1. 

In principle, by further compactifying to lower dimensions and imposing more than one freely acting generator $G_I$ to act on momenta and winding numbers as we have done for $G_5$, it is possible to stabilize more internal moduli, while further decreasing the net value of the potential. However, at low supersymmetry breaking scale, the Scherk--Schwarz direction $X^9$ being large, there cannot be extra massless states charged under the $U(1)$ associated with this coordinate. Therefore, in this regime, the Wilson lines associated with  this Abelian factor cannot be stabilized.


\section{Heterotic \bm $[SO(3)\times SO(1)]^8$ model  in six dimensions}
\label{so31}

As reviewed in Sect.~\ref{Oset}, the moduli space of the  non-perturbatively consistent six-dimensional orientifold models containing only O$5_-$-planes is made of two disconnected branches~\cite{triple}. The first one describes all continuous Wilson line deformations of the usual $SO(32)$ theory, with the rank of the open string gauge group equal to 16. The second one corresponds to the deformations of the orientifold theory with one rigid D5-brane on top of each of the 16 \mbox{O5$_-^\prime$-planes}. In this case, the remaining 16 D5-branes are  free to move in pairs and the rank of the open string gauge group is reduced to 8. The maximal gauge symmetry  is obtained when the 8 pairs of D5-branes are located on top of a single O$5_-^\prime$-plane, yielding $SO(17)\times SO(1)^{15}$. On the other hand, distributing the D5-brane pairs on top of different O$5_-^\prime$-plane reduces the gauge  group to $SO(3)^8\times SO(1)^8$.

In the present section, we show that the orientifold moduli space component of reduced rank admits a heterotic dual description.  For this purpose, it is enough to construct explicitly a dual heterotic model valid at any particular point in moduli space, since its marginal deformations\footnote{At tree level, they are identical whether the Scherk-Schwarz mechanism is implemented or not.}  span all of the moduli space branch.  In the following, we make the choice to realize the dual of an orientifold configuration  with open string gauge group $SO(3)^8\times SO(1)^8$.

In the $SO(1)^{32}$ orientifold model in five dimensions, all five internal directions are on equal footing because of the democratic distribution of the 32 rigid D4-branes on top of the 32 O4-planes. Therefore, it is a matter of convention to implement the Scherk--Schwarz breaking of supersymmetry along any of the five compact directions. However, in the case of an $SO(3)^8\times SO(1)^8$ orientifold configuration in six dimensions, the way of distributing the stacks of D5-branes for a given choice of  Scherk--Schwarz breaking direction does matter.  Indeed, as follows from Eq.~(\ref{cstab}), in order to get a tachyon free model at 1-loop (provided the supersymmetry breaking scale is below the string scale), the gauge symmetry generated by the stacks located on adjacent O5-planes along the Scherk--Schwarz direction must be $SO(3)\times SO(1)$. Using the conventions of Ref.~\cite{ADLP}, the full open string gauge group of such a configuration  is denoted $[SO(3)\times SO(1)]^8$. 

To explicitly construct the dual heterotic model,  an appropriate starting point is the $SO(32)$ heterotic string compactified on $T^4$. In order to reduce the gauge symmetry to $[SO(3)\times SO(1)]^8$, we implement a $\Z_2^4$ orbifold action that realizes the following pattern of breakings
\be
SO(32)\overset{G_9}{\longrightarrow} SO(16)^2\overset{G_8}{\longrightarrow} SO(8)^4\overset{G_7}{\longrightarrow} SO(4)^8\overset{G_6}{\longrightarrow} [SO(3)\times SO(1)]^8\, , 
\ee
where $G_I$, $I\in\{6,\dots,9\}$, denote the $\Z_2^4$ generators. In particular, they act as twists on the extra 32 right-moving worldsheet fermions in the way shown in Table~\ref{T2}.
\begin{table}[h]
\begin{center}
\begin{tabular}{|c|l|}
\hline
$G_9$ & \px\px\px\px\px\px\px\px\px\px\px\px\px\px\px\px\x\x\x\x\x\x\x\x\x\x\x\x\x\x\x\x\\
$G_8$ & \px\px\px\px\px\px\px\px\x\x\x\x\x\x\x\x 	\px\px\px\px\px\px\px\px\x\x\x\x\x\x\x\x \\
$G_7$ & \px\px\px\px\x\x\x\x \px\px\px\px\x\x\x\x \px\px\px\px\x\x\x\x \px\px\px\px\x\x\x\x 	\\
$G_6$ & \px\px\px\x   \px\px\px\x  {\white.}\!\!\!\! \px\px\px\x \px\px\px\x 	{\white.}\!\!\!\! \px\px\px\x   \px\px\px\x {\white .}\!\!\!\!  \px\px\px\x  \px\px\px\x \\
\hline
\end{tabular}
\caption{\label{T2} \em \footnotesize  Twist actions of the four generators $G_I$ on the 32 right-moving real worldsheet fermions. A ``$-$'' sign indicates a non-trivial $\Z_2$ twist of the corresponding fermion.}
\end{center}
\vspace{-0.5cm}
\end{table}
$G_9,G_8,G_7$ are similar to those introduced in Table~\ref{T1}, while $G_6$ now twists only 8 fermions rather than 16.  
In order not to generate massless states in the twisted sectors, which  ensures the $[SO(3)\times SO(1)]^8$ gauge symmetry not to be enhanced back, all the orbifold generators must also act freely on the internal $T^4$ coordinates. Moreover, we choose $G_6$ to be the generator that contains in its definition the additional action of $(-1)^F$, responsible for the spontaneous breaking of supersymmetry~\cite{SSstring-5}.

The 1-loop effective potential 
\be
\Vone=-{1\over (2\pi)^6}\int_{\F} {\d^2 \tau\over 2\tau_2^2} \, Z_{(6)}\, , 
\ee
is expressed in term of the partition function that can be written in the following form,
\begin{align}
Z_{(6)}&={1\over \big(\sqrt{\tau_2}\eta\bar \eta\big)^4}\, {1\over 2^4}\sum_{\vec h,\vec g}e^{i\pi\varphi\left[{}^{\vec h}_{\vec g}\right]}{\Gamma_{4,4}\!\left[{}^{\vec h}_{\vec g}\right]\over (\eta\bar \eta)^4}\, {1\over 2}\sum_{a,b}(-1)^{2(a+b+2ab)}{\theta[{}^a_b]^4\over \eta^4}\, {\Gamma_{0,16}\!\left[{}^{\vec h}_{\vec g}\right]\over \bar \eta^{16}}\, (-1)^{4(g_6a+h_6b+g_6h_6)}\nonumber \\
&\equiv {1\over 2^4}\sum_{\vec h,\vec g}Z_{(6)}\!\left[{}^{\vec h}_{\vec g}\right], 
\label{Z6}
\end{align}
where $\vec h,\vec g$ are 4-vectors, whose components take values 0 or ${1\over 2}$. 
 In these expressions, and accordingly with Table~\ref{T2}, the shifted $(0,16)$-lattice is given by 
\be
\Gamma_{0,16}\!\left[{}^{\vec h}_{\vec g}\right]\!= {1\over 2}\sum_{\gamma,\delta}\bar\theta[{}^\gamma_\delta]^{3\over 2}\, \bar\theta\big[{}^{\gamma+h_6}_{\delta+g_6}\big]^{1\over 2}\, \bar\theta\big[{}^{\gamma+h_7}_{\delta+g_7}\big]^{3\over 2} \cdots \bar\theta\big[{}^{\gamma+h_6+\cdots+h_9}_{\delta+g_6+\cdots+g_9}\big]^{1\over 2}\, ,
\label{16'}
\ee
while the shifted $(4,4)$-lattice and the phase $\varphi\big[{}^{\vec h}_{\vec g}\big]$ we have introduced must be chosen in a way that ensures modular invariance. The latter is guaranteed when the actions of the $SL(2,\Z)$ generators on the Techm\"uller parameter translate into matrix transformations on the conformal blocks,
\begin{alignat}{5}
\label{motrans}
&\tau\to -{1\over \tau} &\Longleftrightarrow  \quad &(\vec h,\vec g)\to (\vec h,\vec g) S\,, \quad (\gamma,\delta)\to (\gamma,\delta)S\, \, ,&  \quad &(a,b)\to (a,b)S\, , \nonumber\\
&{\text{and}}\\
&\tau\to \tau+1  &\Longleftrightarrow  \quad &(\vec h,\vec g)\to (\vec h,\vec g)T\,,\quad  (\gamma,\delta)\to \Big(\gamma,\delta+\gamma-{1\over 2}\Big) ,& \quad &(a,b)\to \Big(a,b+a-{1\over 2}\Big)\nonumber 
\end{alignat}
where
\begin{equation}
S=\left(\!\!\begin{array}{cc} 0 &-1\\ 1 & 0\end{array}\!\!\right) , \qquad T=\left(\!\!\begin{array}{cc} 1 &1\\ 0 & 1\end{array}\!\!\right) ,
\end{equation}
provided  the components of $\vec h,\vec g$ in the definition of $Z_{(6)}\big[{}^{\vec h}_{\vec g}\big]$ can be defined modulo 1.\footnote{$\gamma,\delta$ and $a,b$ appearing in Eqs~(\ref{Z6}) and~(\ref{16'}) are automatically defined modulo 1.}

It turns out that $\Gamma_{0,16}\big[{}^{\vec h}_{\vec g}\big]$ is invariant under either of the shifts $h_I\to h_I+1$ or $g_I\to g_I+1$, $I\in\{6,\dots,9\}$, but only satisfies the transformations~(\ref{motrans}) up to non-trivial multiplicative phases. To compensate these phases, we may choose 
\be
\varphi\!\left[{}^{\vec h}_{\vec g}\right]\!= 2\sum_{I=6}^9 (h_6 g_I+g_6h_I)\, , 
\ee
so that the product $e^{i\pi\varphi\left[{}^{\vec h}_{\vec g}\right]}\Gamma_{0,16}\big[{}^{\vec h}_{\vec g}\big]$ indeed satisfies the transformations rules~(\ref{motrans}). However, the price to pay once we introduce the non-trivial phase $\varphi\big[{}^{\vec h}_{\vec g}\big]$  is that $e^{i\pi\varphi\left[{}^{\vec h}_{\vec g}\right]}\Gamma_{0,16}\big[{}^{\vec h}_{\vec g}\big]$ is no more invariant under the shifts $h_I\to h_I+1$ and $g_I\to g_I+1$. Fortunately,  the torus lattice $\Gamma_{4,4}\big[{}^{\vec h}_{\vec g}\big]$ has similar properties, when the orbifold generators act asymmetrically on the internal coordinates. Namely, it respects the transformation rules~(\ref{motrans}), but is not invariant under the shifts  $h_I\to h_I+1$ and $g_I\to g_I+1$, as explained  below Eq.~(\ref{pg}). Using this observation, it appears that the following suitable choice of $(4,4)$-lattice, 
\be
\begin{aligned}
&\Gamma_{4,4}\!\left[{}^{\vec h}_{\vec g}\right]\!= \Gamma_{4,4}\!\left[{}^{\vec \goh,\vec \goh'}_{\vec \gog,\vec \gog'}\!\right],&& \espD\\
\where \quad &\goh_I=h_I\, , \;\; I\in\{6,7,8,9\}\, , \quad &&\goh'_6=h_7+h_8+h_9\, ,\quad &&\goh'_7=\goh'_8=\goh'_9=h_6\, ,\\
&\gog_I=g_I\, ,  \quad &&\gog'_6=g_7+g_8+g_9\, ,\quad &&\gog'_7=\gog'_8=\gog'_9=g_6\, ,\\
\end{aligned}
\ee
implies all shifts $h_I\to h_I+1$ and  $g_I\to g_I+1$ to become symmetries of $e^{i\pi\varphi\left[{}^{\vec h}_{\vec g}\right]}\Gamma_{4,4}\big[{}^{\vec h}_{\vec g}\big]\Gamma_{0,16}\big[{}^{\vec h}_{\vec g}\big]$. In that case, the  full partition function in Eq.~(\ref{Z6}) is modular invariant. To put it plainly, in the internal space, $G_6$ acts as a half-period shift on $X_L^6+X_R^6$, as well as on $X_L^7-X^7_R$,  $X_L^8-X^8_R$ and $X_L^9-X^9_R$, while each of the remaining generators $G_I$, $I\in\{7,8,9\}$, acts as a half-period shift on $X_L^I+X^I_R$, as well as on $X_L^6-X_R^6$. Because all $G_I$'s shift a geometric coordinate $X^I=X_L^I+X_R^I$, their actions are  free, as  required. 
In total, the different blocks of the partition function take the Hamiltonian form
\be
\begin{aligned}
Z_{(6)}\!\left[{}^{\vec h}_{\vec g}\right]\!= & \;{(-1)^{2\varphi\left[{}^{\vec h}_{\vec g}\right] +4h_6g_6}\over \tau_2^2(\eta\bar \eta)^8}\,  {1\over 2}\sum_{a,b}(-1)^{2(a+b+2ab)}\,{\theta[{}^a_b]^4\over \eta^4}\, (-1)^{4(g_6a+h_6b+g_6h_6)}\, {\Gamma_{0,16}\!\left[{}^{\vec h}_{\vec g}\right]\over \bar \eta^{16}}\\
&\;\sum_{\vec m,\vec n}    (-1)^{2[g_6(m_6+n_7+n_8+n_9)+g_7(m_7+n_6)+g_8(m_8+n_6)+g_9(m_9+n_6)]}\, \Lambda_{\vec m+\vec  \goh', \vec n+\vec h}\,.
\end{aligned}
\ee 

In order to describe the spectrum,  it is useful to write  the above result in terms of affine $O(2n)$ characters. In the untwisted sector $\vec h=\vec 0$, because the  generator $G_6$ twists 8 real fermions, we obtain 
\be
\begin{aligned}
Z_{(6)}\!\left[{}^{\vec 0}_{(g_6,0,0,0)}\right]\!={1\over \tau_2^2(\eta\bar\eta)^8}\,& \Big(V_8-(-1)^{g_6}S_8\Big) \sum_{\vec m,\vec n}(-1)^{2g_6(m_6+n_7+n_8+n_9)}\Lambda_{\vec m,\vec n}\\
&\Big( \bar O_{24}\bar O_{8}-(-1)^{\delta_{g_6,0}}\bar V_{24}\bar V_{8}+\bar S_{24}\bar S_{8}-(-1)^{\delta_{g_6,0}}\bar C_{24}\bar C_{8}\Big).
\end{aligned}
\label{Zhg'}
\ee
Moreover, one can see from Table~\ref{T2} that the remaining 14 group elements $\prod_I G_I^{2g_I}$, where $(g_7,g_8,g_9)\neq (0,0,0)$, twist 16 real fermions, leading to
\begin{align}
Z_{(6)}\!\left[{}^{\vec 0}_{\vec g}\right]\!={1\over \tau_2^2(\eta\bar\eta)^8}& \Big(V_8-(-1)^{g_6}S_8\Big)\! \!\sum_{\vec m,\vec n} (-1)^{2[g_6(m_6+n_7+n_8+n_9)+g_7(m_7+n_6)+g_8(m_8+n_6)+g_9(m_9+n_6)]} \Lambda_{\vec m, \vec n}\nonumber \\
&\Big( \bar O_{16}\bar O_{16}-\bar V_{16}\bar V_{16}+\bar S_{16}\bar S_{16}-\bar C_{16}\bar C_{16}\Big).
\label{Zhg''}
\end{align}
Similarly, in the sector twisted only by $G_6$, we have 
\be
Z_{(6)}\!\left[{}^{(1,0,0,0)}_{\vec 0}\right]\!={1\over \tau_2^2(\eta\bar\eta)^8}\, \Big(O_8-C_8\Big)\sum_{\vec m,\vec n}\Lambda_{\vec m+\vec \goh',\vec n+\vec h}\Big( \bar O_{24}\bar S_{8}+\bar V_{24}\bar C_{8}+\bar S_{24}\bar O_{8}+\bar C_{24}\bar V_{8}\Big),
\label{ztw}
\ee
while for the remaining 14 twisted sectors $\vec h$, where $(h_6,h_7,h_8)\neq (0,0,0)$, we find
\be
\begin{aligned}
Z_{(6)}\!\left[{}^{\vec h}_{\vec 0}\right]\!={1\over \tau_2^2(\eta\bar\eta)^8}\,& \!\Big[\delta_{h_6,0}\big(V_8-S_8\big) +\delta_{h_6,{1\over 2}}\big(O_8-C_8\big)\Big]\sum_{\vec m,\vec n}\Lambda_{\vec m+\vec \goh',\vec n+\vec h}\\
&\Big( \bar O_{16}\bar S_{16}+\bar V_{16}\bar C_{16}+\bar S_{16}\bar O_{16}+\bar C_{16}\bar V_{16}\Big).
\end{aligned}
\label{ztw'}
\ee

Due to the free action of the generator $G_6$  on $X^6$, it is not surprising that all the states involved in Eq.~(\ref{ztw}) are automatically ``super-heavy'', when the scale  
\be
M={\sqrt{G^{66}}\over 2} \, ,
\ee
of supersymmetry breaking is lower than the string scale (the winding number $n_6+{1\over 2}$ cannot vanish).\footnote{Tachyons in $O_8\bar O_{24}\bar S_{8}/(\eta\bar \eta)^8$, and  massless bosons and fermions in $O_8\bar V_{24}\bar C_{8}/(\eta\bar \eta)^8$ and $-C_8\bar O_{24}\bar S_{8}/(\eta\bar \eta)^8$ may arise, when $M=\O(1)$.} Similarly, the free actions of $G_7,G_8,G_9$ respectively on the directions $X^7,X^8,X^9$  imply all the states in the sectors $(h_6,h_7,h_8)\neq (0,0,0)$ to also be very heavy, when the sizes of these directions are not much smaller than the string length (the vector of winding numbers $\vec n+\vec h$ cannot vanish).
Therefore, when $M\ll 1$ and the other compactification moduli are generic, the light spectrum, as anticipated,  is composed of the  massless states of the untwisted sector, as well as their towers of modes associated with the large internal direction $X^6$. The massless bosons arise in the combinations of characters 
 $V_8\bar O_{24}\bar O_{8}/(\eta\bar\eta)^8$, $V_8 \bar V_{24}\bar V_{8}/(\eta\bar\eta)^8$ in Eq.~(\ref{Zhg'}), and  $V_8\bar O_{16}\bar O_{16}/(\eta\bar\eta)^8$, $V_8 \bar V_{16}\bar V_{16}/(\eta\bar\eta)^8$ in Eq.~(\ref{Zhg''}).  Their counting goes as follows, 
\be
\begin{aligned}
n_B={8\over 2^4}\Big(& \big[8+276+28+24\times 8\big]+\big[8+276+28-24\times 8\big]\\
&+14 \times \big[8+120+120-16\times 16\big]\Big)= 8\times (8+8\times 3)\, ,
\end{aligned}
\ee
which matches with the spectrum expected from the orientifold point of view, Eq.~(\ref{nfbtotD'}). In total, the gauge group symmetry is $U(1)_{\rm grav}^4\times U(1)^4\times [SO(3)\times SO(1)]^8$. Similarly, the massless fermions are found from the characters $-S_8\bar O_{24}\bar O_{8}/(\eta\bar\eta)^8$, $-S_8 \bar V_{24}\bar V_{8}/(\eta\bar\eta)^8$ in Eq.~(\ref{Zhg'}), and $-S_8\bar O_{16}\bar O_{16}/(\eta\bar\eta)^8$, $-S_8 \bar V_{16}\bar V_{16}/(\eta\bar\eta)^8$ in Eq.~(\ref{Zhg''}). Notice that the sectors $g_6={1\over 2}$ contribute with an opposite sign, as follows from the fact that the generator $G_6$ also acts as $(-1)^F$. As a result, we find 
\be
\begin{aligned}
n_F={8\over 2^4}\Big( &\big[8+276+28+24\times 8\big]-\big[8+276+28-24\times 8\big]\\
&+(7-7) \times \big[8+120+120-16\times 16\big]\Big)= 8\times (8\times 3)\, ,
\end{aligned}
\ee
again reproducing the counting of the fermionic degrees of freedom found in 8 copies of vector multiplets in the bifundamental representation of $[SO(3)\times SO(1)]$. 

To summarize, the heterotic model we have constructed fulfils all the requests a dual version of the orientifold models under consideration should satisfy. Notice that, on the open string side, several distributions of D5-branes yield an $[SO(3)\times SO(1)]^8$ gauge symmetry, with massless fermions in the bifundamentals. They are obtained by exchanging the numbers of D5-branes located on a pair of adjacent O5-planes along the Scherk--Schwarz direction. Even though this operation does not change the light spectrum, it does change the perturbative heavy modes. However, as can be seen in Table~\ref{T2}, the 32 fermions of the heterotic description can be split into 8 equivalent sets of 4. Because the generator $G_6$ flips the sign of any arbitrary fermion out of 4 in each set, it seems that the heterotic description is unique. Thus, all the ``cousin'' orientifold configurations described above should be  equivalent to each-other  at the non-perturbative level.


\section{Conclusions}
\label{cl}

The fact that branes may have rigid positions in orientifold models entails appealing possibilities for reducing moduli instabilities that generically occur at the quantum level, when supersymmetry is spontaneously broken. They can also be used to increase  the 1-loop potential to a vanishing or positive value, at least in a perturbative framework. However, strong constraints arising from the consistency of the setup at the non-perturbative level imply a drastic reduction of the allowed configurations.  In particular, the number of rigid branes can only be  0, 16 or 32, with specific distributions on the O$_-^\prime$-planes. In the present work, we have shown that all non-perturbatively-consistent branches of the  toroidal orientifold moduli space in dimension $D\ge 5$ admit heterotic dual descriptions. This is done by considering the usual $SO(32)$ heterotic theory, and by explicitly constructing two representative models of reduced rank, with gauge groups $[SO(3)\times SO(1)]^8$ and $SO(1)^{32}$. 

The $SO(32)$ and $[SO(3)\times SO(1)]^8$ models have negative potentials, with respectively massive or marginal Wilson lines at 1-loop, provided that the supersymmetry breaking scale is below the string scale to avoid severe tree level instabilities similar to the Hagedorn transition~\cite{Atick,AK,ADK,PV1,PV2}. On the contrary, the $SO(1)^{32}$ theory does not develop tree-level instability of this type at any point in moduli space, while its potential is positive at least at low supersymmetry breaking scale.\footnote{It can be checked that in each branch, the sign of $n_F-n_B$ is common to all dual pairs that are tachyon free at 1-loop.}

An interesting fact is that the heterotic counterpart of rigid D-branes is realized by freely acting orbifolds that assign to internal real free fermions boundary conditions in such a way that one is left with two-dimensional Ising model conformal blocks. 

Unfortunately, the issue of moduli stabilization is more subtle than expected. In the open string side, the closed string moduli are flat directions due to the lack of  light states\footnote{This is not true for the heavy states but in the regime studied throughout the present paper, their contributions to the 1-loop potential are exponentially suppressed, $\O(e^{- {2\pi\over M}})$.} charged under the Abelian symmetries arising from the toroidal compactification. One interest into looking for heterotic duals lied in the fact that {\it a priori} these scalars may  be stabilized in a dynamical way at points in moduli space of enhanced gauge symmetry. However, the symmetric or asymmetric orbifold actions constrained by modular invariance tipically project out such extra massless states, jeopardizing the lifting of the corresponding flat directions.  Moreover, we only get models with unbalanced bosonic {\it versus} fermionic massless degrees of freedom, implying the effective potential to be large, due to its scaling proportional to a power of the supersymmetry breaking scale.


 \section*{Acknowledgements}

The authors are grateful to Steven Abel, Ignatios Antoniadis, Emilian Dudas, Kumar Narain and especially Daniel Lewis for useful inputs during the realization of this work. 
The work of C.A. is partly supported by the MIUR-PRIN contract 2017CC72MK\_003.
The work of H.P. is partially supported by the Royal-Society/CNRS International Cost Share Award IE160590. 
C.A. would like to acknowledge the hospitality of CPHT of the Ecole Polytechnique and of the Theory Department at CERN where part of this work has been done.
H.P. would like to thank the CERN Theory Department and the Physics Department of the University of Rome Tor Vergata, where part of this work has been performed. G.P. would like to thank the CPHT of the Ecole Polytechnique, where  part of this work has also been developed.


\newpage 

 \bibliographystyle{unsrt}

\end{document}